\documentstyle[aps,prd]{revtex}
\input axodraw1.sty
\def\be{\begin{equation}}
\def\ee{\end{equation}}
\def\beq{\begin{eqnarray}}
\def\eeq{\end{eqnarray}}
\def\s{\sigma}

\def\G{\Gamma}

\def\an{analytic}
\def\ac{\an{} continuation}
\def\hsr{hypergeometric series representations}
\def\hf{hypergeometric function}
\def\ndim{NDIM}

\def\lra{\leftrightarrow}
\begin{document}

\draft
\title{Non-planar double-box, massive and massless pentabox Feynman
integrals in negative dimensional approach}
\author{Alfredo T. Suzuki\footnote{E-mail:suzuki@ift.unesp.br}, Alexandre G. M.
Schmidt\footnote{E-mail:schmidt@ift.unesp.br}}
\address{Universidade Estadual Paulista --- Instituto de F\'{\i}sica Te\'orica
R.Pamplona, 145 S\~ao Paulo - SP CEP 01405-900, Brazil}
\date{\today}\maketitle

\begin{abstract}

Negative dimensional integration method (NDIM) is a technique
which can be applied, with success, in usual covariant gauge
calculations. We consider three two-loop diagrams: the scalar
massless non-planar double-box with six propagators and the scalar
pentabox in two cases, where six virtual particles have the same
mass and in the case where all of them are massless.  Our results
are given in terms hypergeometric functions of Mandelstam
variables and for arbitrary exponents of propagators and dimension
$D$ as well.
\end{abstract}


\section{Introduction}

Studies in particle phenomenology require more and more sophisticated
calculations\cite{gehrmann}, and the measurement of the $(g-2)$ factor
has now an error of $1\:ppb$ order of magnitude \cite{kinoshita}
thanks to the perturbative approach. Within this perspective there
have been some interest in massless double-box --- planar and
non-planar\cite{oleari} --- and pentabox integrals.
Smirnov\cite{smirnov} studied the former, using the Mellin-Barnes (MB)
technique, in three cases: scalar and tensorial with four legs
on-shell and scalar with one leg off-shell; Tausk\cite{tausk} {\bf and
Smirnov and Veretin\cite{veretin}}, also using the MB method,
presented {\bf explicit} results for non-planar, or crossed,
double-box with seven and six propagators, in the special case where
the exponents of propagators are equal to one; K\"ummer {\it et al}
came across with the same integrals studying the potential between
quarks in the Coulomb gauge\cite{kummer}; Anastasiou {\it et al}
calculated the latter, in the integration-by-parts approach, for both
the scalar and tensorial cases with massless internal
particles\cite{pentabox}. The results which are obtained with the MB
approach are, like the ones obtained by \ndim{}, expressed in terms of
infinite series of hypergeometric type. Of course, progress along this
line is greater in covariant gauges, however perturbative calculations
in non-covariant ones are also carried out and need sometimes more
powerful techniques than the formers\cite{bassetto}.

In this paper we choose to tackle the scalar on-shell non-planar
double-box integral for arbitrary exponents of propagators, a
result that is missing in the literature and the pentabox integral
in two cases: where all particles are massless, and where six
virtual particles have the same mass, a diagram which, for
instance, contributes to photon-photon scattering, and as far as
we know was not yet calculated before. \ndim{} gives us several
results in terms of Mandelstam variables and masses, each valid in
a certain kinematical region. We give results in terms of
hypergeometric series for arbitrary exponents of propagators and
dimension. In our approach no reduction formulas or
integration-by-parts methods are used or even necessary. It is
also worth observing that \ndim{} is a technique which can be
applied to other gauge choices like the Coulomb and the light-cone
gauges \cite{outstanding}.

An important feature of \ndim{} is that {\it it is not } a
regularization technique. It is worth remembering Dunne and
Halliday\cite{dunne}: the negative-dimensional integrals (in
$D$-dimensions) can be related to positive dimensional ones (in
$2N$-dimensions) over Grassmannian variables; in fact, one has just to
make $D\leftrightarrow -2N$. So, in \ndim{} context there are no
singularities, no poles etc. However, when we perform the \ac{}, in
order to allow negative exponents of propagators and positive
dimension, then poles appear for specific values of those exponents
and physical $D=4$ dimension and we have the same results which other
techniques provide. This is therefore a consistent method to solve
Feynman loop integrals pertaining to usual covariant or to
non-covariant algebraic gauges, like Coulomb and light-cone ones (even
at the two-loop level).

The aim of our paper is not to establish the axiomatic foundations for
NDIM nor to demonstrate in a rigorous mathematical sense its
principles and basis.  Instead, given the simple steps that the
methodology requires to work out complicated Feynman integrals we are
interested in testing it to the limits of our present calculational
abilities. For this purpose we are presenting here in
another\cite{2-3loops} true two-loop calculation the exact results
yielded by this method. Such results must be compared with others
obtained using different techniques so that not only previous answers
be double-checked and the confidence in the novel method be increased
but also in order to demonstrate the feasibility of the latter in
performing the calculations.

The outline for our paper is as follows: in the next section we study
pentabox integrals, where in one case virtual particles are massless
and in the other case, we consider six of the virtual particles
massive with equal masses, a graph which contributes to photon-photon
scattering (which were considered recently in \cite{bern}), although a
full calculation of such an effect as well as evaluation of beta
functions in physical processes are beyond the scope of the present
work.  We solve also the scalar massless non-planar double-box with
six propagators. In section 3, we present our conclusions.

\section{Pentabox and non-planar double-box integrals}

Let us define immediately the relevant three negative-dimensional
integrals, namely,

\be\label{pentabox}
{\cal P} = \int \int d^D\! q\; d^D\! k_1\; {\bf P}(q,k_1,p,p',p_1) ,
\ee
\be\label{penta-mass}
{\cal MP} = \int \int d^D\! q\; d^D\!k_1\; {\bf
MP}(q,k_1,p,p',p_1,\mu) ,
\ee
and
\be\label{np6}
{\cal NP}_6 = \int \int  d^D\! q\; d^D\! r\; {\bf NP_6}(q,r,p_2,p_3,p_4)\;,
\ee
where the integrands are respectively
\be
{\bf P}\equiv (q^2)^i
(q-p)^{2j} (q-k_1)^{2k} (q-k_1-p_1)^{2l} (k_1^2)^m  (q-p_1)^{2n}
(q-p-p')^{2r}  , \nonumber
\ee
\beq
{\bf MP}&\equiv &(q^2-\mu^2)^i \left[(q-p)^2 -\mu^2\right]^j
\left[(q-k_1)^2-\mu^2\right]^k \left[(q-k_1-p_1)^2-\mu^2\right]^l
(k_1^2)^m \\ \nonumber
&\times &\left[(q-p_1)^2-\mu^2\right]^n \left[(q-p-p')^2-\mu^2\right]^r\; ,
\nonumber
\eeq
\be
{\bf NP_6}\equiv (q^2)^i
(q-p_3)^{2j} (q+r)^{2k} (q+r+p_2)^{2l} (r^2)^m  (r-p_4)^{2n}\;, \nonumber
\ee
which represent massless pentabox, massive pentabox and non-planar
double-box, respectively. Once they are introduced, we evaluate
them using \ndim{}.

When Halliday and co-workers advanced the idea of negative-dimensional
integration they proved that it is equivalent\cite{dunne} to
Grassmannian integration in positive dimension, the correspondence
being as simple as $D\lra -2N$. This fact is implied in the very
structure of the above integrands.

\subsection{Massless Pentabox integral}

\begin{figure} \label{pentabox-fig}
\begin{center}
\begin{picture}(600,200)(0,150)
\vspace{40mm} \thinlines
\put(20,200){\line(1,0){300}}  
\put(20,190){\vector(1,0){30}}\put(20,300){\line(1,0){300}} 
\put(20,310){\vector(1,0){30}}
\put(50,200){\line(0,1){100}} 
\put(280,300){\line(0,-1){100}} \put(280,240){\line(-1,1){60}}
\put(290,190){\vector(1,0){30}} \put(290,310){\vector(1,0){30}}
{\small \put(170,180){\makebox(0,0)[b]{$q-p-p'$}}
\put(310,320){\makebox(0,0)[b]{$p_1$}}
\put(310,180){\makebox(0,0)[b]{$p+p'-p_1$}}
\put(10,320){\makebox(0,0)[b]{$p$}}
\put(10,180){\makebox(0,0)[b]{$p'$}}
\put(140,310){\makebox(0,0)[b]{$q$}}
\put(300,220){\makebox(0,0)[b]{$q-p_1$}}
\put(70,245){\makebox(0,0)[b]{$q-p$}}
\put(255,310){\makebox(0,0)[b]{$q-r'$}}
\put(310,270){\makebox(0,0)[b]{$q-r'-p_1$}}
\put(235,260){\makebox(0,0)[b]{$r'$}}}
\end{picture}\caption{Scalar massless pentabox with all external legs on-shell. Mandelstam variables are defined
as $s=(p+p')^2$ and $t=(p-p_1)^2 $}
\end{center}\end{figure}
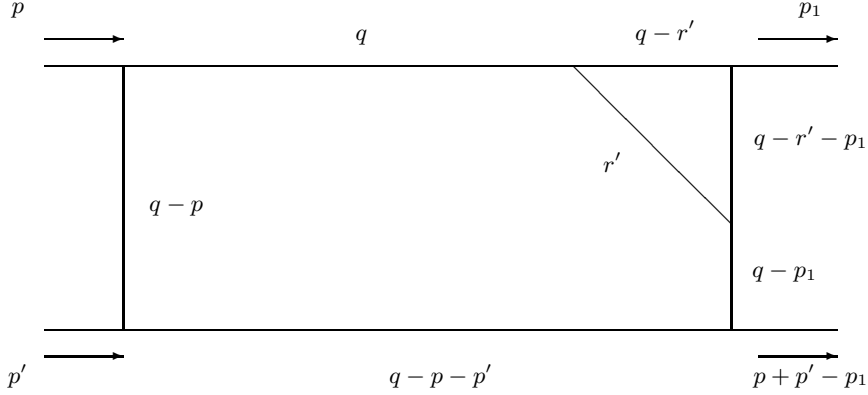

The negative-dimensional integral for massless pentabox diagram ---
see Figure 1 --- has the following generating functional,

\beq\label{geradora-P} G_{{\cal P}} &=& \int \int d^D\! q\; d^D\! r'\;
\exp{\left[-\alpha q^2 -\beta (q-p)^2 -\gamma (q-r')^2 -\theta
(q-r'-p_1)^2 -\phi r'^2 -\eta (q-p_1)^2 -\omega (q-p-p')^2\right]}
\\  &=& \left(\frac{\pi^2}{\Lambda}\right)^{D/2}
\exp{\left[-\frac{1}{\Lambda}\left( \alpha \lambda_2\omega s +
\beta\eta\lambda_2 t + \beta\theta\phi t \right)\right] }
,\label{ger-P}
\eeq where $\lambda_2 = \gamma + \theta + \phi$ and
$\Lambda = (\alpha +\beta +\eta +\omega )\lambda_2 + \gamma\phi +
\theta\phi$.

Taylor expanding (\ref{geradora-P}) we have our
negative-dimensional integral ${\cal P}$ as a factor in a seven-fold
summation series,
\be\label{projetando} G_{{\cal P}}
= \sum_{i,j,k,l, m,n,r=0}^\infty \frac{ (-1)^{i+j+k+l+m+n+r}}{i!
j!k!l!m!n!r!}\alpha^i \beta^j\gamma^k \theta^l\phi^m\eta^n\omega^r
\; {\cal P}. \ee

On the other hand taking (\ref{ger-P}) and
making an expansion (including a multinomial one) in power series, we
obtain,
\beq G_{{\cal P}} &=&
\sum_{X,Y,Z=0}^\infty \frac{\alpha^{X_{123}+Y_{123}}
\beta^{X_{4567}+Y_{456}} \gamma^{X_{14}+Y_{147}+ Z_{13}}
\theta^{X_{257}+ Y_{258}+Z_{24}} \phi^{X_{367}+ Y_{369}+
Z_{125}}\eta^{X_{456}+ Y_{789}} \omega^{X_{123}+Z_{345}} }{
X_1!...X_7!Y_1!...Y_9!Z_1!...Z_5!}\nonumber\\
&&\times (-s)^{X_{123}}(-t)^{X_{4567}}(-X_{1234567}-D/2)!, \eeq
where sum over $X, Y, Z$ is a shorthand notation for a 7-, 9-, and
5-fold sums respectively. Moreover, we define $ X_{12} = X_1 +
X_2,... $, so on and so forth.

Therefore, the exponential above generates a series indexed by 21
indices, while the 7 propagators in the argument of the integrand give
rise to 7 equations and the multinomial expansion to another one, see
eq.(\ref{projetando}). Now, solving both equations for ${\cal P}$ we
conclude that there must be some relations among the two sets of
indices $\{X,Y,Z\}$ and $\{i,j,k,l,m,n,r\}$. It is a system of
algebraic equations,
\be\label{sistema} \left\{ \matrix{ X_{123} +
Y_{123} &=& i \cr
                     X_{4567} +Y_{456} &=& j \cr
                     X_{14}+Y_{147}+Z_{13} &=& k \cr
                     X_{257}+Y_{258}+Z_{24}&=& l \cr
                     X_{367}+Y_{369}+Z_{125} &=& m \cr
                     X_{456}+Y_{789}+Z_{13} &=& n \cr
                     X_{123}+Z_{345} &=& r \cr
                     \Sigma X+\Sigma Y + \Sigma Z &=& - D/2 }\right. ,\ee
for which there is not a unique way to solving it, since we have 21
``unknowns'' and just 8 equations. In fact, there are 203,490 possible
$8\times 8$ systems which can be solved in terms of exponents of
propagators $i,j,k,l,m,n,r$, dimension $D$ and some of $X,Y,Z$. The
computer can calculate such $ 8\times 8$ determinants easily: 134,890
of them are zero, i.e, give empty sets of solutions.

So, our result will be written in terms of a 13-fold series of
hypergeometric type. This can be worked out conveniently and be
simplified, since we have three possible variables: $t/s$, $s/t$, and
unity. Series which depend on both $t/s$ and $s/t$ can not be
convergent, so we will not consider them.

Next, our strategy is to search for the simplest hypergeometric series
among the remaining $68,600$ solutions. The criterion for this search
is dictated by the fact that the more sums with unity argument the
simpler it is, since one can sum them (at least in principle,
respected some relations amongst the parameters) using Gauss' summation
formula\cite{luke},
\be _2F_1(a,b;c|1) =
\frac{\G(c)\G(c-a-b)}{\G(c-a)\G(c-b)}, \;\; {\rm Re}(c-a-b)>0. \label{soma}
\ee

In other words, when some of those 13 sums can be rewritten as gamma
functions. Of course, not all sums with unity argument can be summed
this way, as we will shortly see. These cases occur when the resulting
function is of the type $_3F_2(...|1)$, or even more complex ones.

There are many different ways in which the 13-fold series appear, and
we can classify them according to the following general form,
$$
{\rm (Momenta) (Gammas)} \sum_{\rm all=0}^\infty
\frac{z^A}{n_1!...n_9!m_1!...m_4!}, $$
where $z$ is any one of the three possible variables $t/s$, $s/t$, or
$1$, and $A$ represents either $n_{123}=n_1+n_2+n_3$, or
$n_{1234}=\sum_{i=1}^4n_i$, or $n_{12345}=\sum_{i=1}^5n_i$.

The simplest \hsr{} for the scalar massless pentabox is given by
hypergeometric series with three such ``variables'' (here we use the
word ``variables'', to denote the one variable $s/t$ appearing thrice
as summation variables within the series):

\be\label{penta-3som}
{\cal P}_3 = \pi^D\,s^{\s-j}\,t^j\, P_3^{AC}\:{\bf \Sigma(Y_4,Y_5,Y_6)}
\, {}_3F_2(a_3,b_3,c_3;e_3,f_3|1)
\ee
where
\beq
{\bf \Sigma(Y_4,Y_5,Y_6)}&\equiv &\sum_{Y_{4},Y_{5},Y_6=0}^\infty
\frac {(-k|Y_4)(j-k-l-m-n-D/2|-Y_{456})(D+k+l+m+n+r|Y_{456})}
{(1-k-m-D/2|-Y_5)(D+k+l+m|Y_{456})}\\ \nonumber
&\times& \frac {(D+i+k+l+m+n|Y_{456})(k+l+D/2|Y_6)(-j|Y_{456})}
{(1+\s-j|Y_{456})}\frac{(s/t)^{Y_{456}}}{Y_4!\,Y_5!\,Y_6!}\,,
\eeq
and the parameters of $_3F_2$ are
\beq
a_3&=&-k+Y_4,\\ \nonumber
b_3&=&j-k-l-m-n-D/2-Y_{456},\\ \nonumber
c_3&=&-k-l-m-D/2,\\ \nonumber
e_3&=&1-k-m-D/2-Y_5,\\ \nonumber
f_3&=&-k-l-m-n-D/2 \nonumber
\eeq
with $\s=i+j+k+l+m+n+r+D$ and
\beq
P_3^{AC} &=& (-i|\s-j)(-k-l-m-n-D/2|j)(-l|k+l+m+D/2)(-m|-k-l-D/2) \nonumber\\
&\times&(-r|k+l+r+D/2)(\s+D/2|-2\s-D/2+j)(k+l+m+D|i+n)\,
\eeq

In the above equations, we have used the subscript ``3'' and
superscript ``AC'' to mean that we have three ``variables'' and the result
is analitically continued into positive dimension $D$. Pochhammer
symbols and some of their properties are used throughout the
expressions,
\be (a|b) \equiv
(a)_b = \frac{\G(a+b)}{\G(a)}, \qquad (a|-k) =
\frac{(-1)^k}{(1-a|k)}, \qquad (a|b+c) = (a+b|c) (a|b). \ee

Observe that in (\ref{penta-3som}) there is a fourth sum, namely in
$Z_1$, which has unity argument and could not be summed because it is
a ${}_3F_2$ hypergeometric series which can be expressed in terms of
gamma functions only in some special cases. This means that we were
able to sum up up to 9 series (with unity arguments) using (\ref{soma}).

The second type of result provided by our method is given by
hypergeometric series with four ``variables'',

\beq \label{penta-4som}
{\cal P}_4 &=& \pi^D\,s^{\s}\, P_4^{AC}\;
\sum_{X_n=0}^\infty
\frac{(-j|X_{4567})(-n|X_{456})(-k|X_4)(-l|X_{57})(-\s|X_{4567})}
{(i-\s|X_{4567})(r-\s|X_{4567})(-k-l|X_{457})} \nonumber \\ &\times &
\frac{(D/2+m|X_{45})(D/2+k+l|X_6)(-k-l-m-D/2|X_7)}{(D+k+l+m|X_{456})}
\frac{(t/s)^{X_{4567}}}{X_4!X_5!X_6!X_7!}\,,
\eeq
where the subscript ``4'' means a 4-fold \hsr{} for ${\cal P}$, and
\be
P_4^{AC} = (-i|\s) (-r|\s) (-k-l|-m-D/2) (\s+D/2|-2\s-D/2)
(k+l+m+D|-m-D/2) (-m|2m+D/2).
\ee

Note that this four-fold series can be used to study forward
scattering. Taking $t=0$ we are left with only the first term in
the series, which is equal to unity, that is, the series collapses
and the result is merely (the superscript FS for forward
scattering case)

$${\cal P}_4^{[FS]} = \pi^D\;s^{\s}\;P_4^{AC}.$$

The last one, given by hypergeometric series with five
``variables'',

\beq \label{penta-5som}
{\cal P}_5 &=& \pi^D\;s^{i+j+r+D/2}\;t^{k+l+m+n+D/2}\:P_5^{AC} \nonumber \\
&\times & \sum_{Y_n,Z_m=0}^\infty
\frac{(-n|Y_{789})(-k|Y_7+Z_1)(-k-l-m-D/2|Z_{12})(k+l+D/2|Y_9)
}{(1-k-m-D/2|Y_{789})(1+j-k-l-m-n-D/2|Y_{789}+Z_{12})(D+k+l+m|Y_{789})}
\nonumber \\
&\times & \frac{(j+r+D/2|Y_{789}+Z_{12})(i+j+D/2|Y_{789}+Z_{12})}{(1+i+j+r+D/2|Y_{789}+Z_{12})}\,\frac{(s/t)^{Y_{789}+Z_{12}}}{Y_7!Y_8!Y_9!Z_1!Z_2!}\,,
\eeq
where
\beq P_5^{AC} &=&
(-i|-j-r-D/2)(-j|k+l+m+n+D/2)(k+l+m+D|-m-D/2)(-l|-k-m-D/2) \nonumber\\
&\times &(-m|i+j+m+D/2) (-r|k+m+r+D/2) (\s+D/2|j+r-\s). \eeq

Observe that in the last line of (\ref{penta-5som}) when we take
either $i=-1$ or $r=-1$ those pertinent Pochhammer symbols within the
series do cancel out, simplifying it.

Of course, there are results which depend on the inverse of such
variables, i.e., ${\cal P}_3(t/s)$, ${\cal P}_4(s/t) $ and ${\cal
P}_5(t/s)$, which means an interchange of one or more pairs of
exponents of propagators and $s\lra t$.

How are these different expressions related to one another? How do
they relate to previous calculations by other methods, for example
for special cases of the propagator powers? How unique or
ambiguous is the analytic continuation in dimension to get back to
answers in positive dimensions? What form of renormalization is
necessary to connect with conventional calculations?

To answer the first question one must recall that if two series
(${\cal P}_3$ and ${\cal P}_5$, for instance) represent the same
function (the integral ${\cal P}$) then they must be related by
\ac{} \cite{feshbach}. So, the results provided by \ndim{} are
always related by \ac{} either directly (overlapping regions of
convergence) or indirectly (no overlapping of regions). In the
present case there are no previous calculations in the literature
considering arbitrary exponents of propagators.

\subsection{Massive Pentabox integral}

Introducing masses in the \ndim{} context is very
simple\cite{ach}. The generating functional becomes,

\beq
G_{MP} &=& \int \int d^D\! q\; d^D\! k_1\; \exp{\left\{-\alpha
(q^2-\mu^2)
-\beta [(q-p)^2-\mu^2]-\gamma [(q-k_1)^2-\mu^2] -\theta [(q-k_1-p_1)^2-\mu^2] \right.}\\
&& \left. -\phi k_1^2 -\eta [(q-p_1)^2-\mu^2]- \omega
[(q-p-p')^2-\mu^2]\right\}
\nonumber\\
&=& \left(\frac{\pi^2}{\Lambda}\right)^{D/2}
\exp{\left[-\frac{1}{\Lambda}\left( \alpha \lambda_2\omega s +
\beta\eta\lambda_2 t + \beta\theta\phi t \right)\right] }
\exp{\left[ \left(\alpha +\beta
+\gamma+\theta+\eta+\omega\right)\mu^2\right] },
\eeq
so that we have six additional sums, which are generated by the second
exponential in the second line above, which corresponds to the massive
sector. Besides, the resulting hypergeometric series will have
variables of the form,
$$ \frac{t}{s}, \qquad\frac{t}{\mu^2}, \qquad\frac{s}{\mu^2}, $$ their
inverses or unity. Powers of them also occur, like $\sqrt{s/t}$ and
$(s/t )^2$.

Once more we look for convergent series. Among the very BIG number of
possible systems --- altogether $27!/(8!19!)=2,220,075$ --- there are
$1,093,289$ which have no solution, and the remaining $1,126,786$
among which we are able to find solutions. So we are left with 49.24\%
of the total, from which we hunt for the most convenient ones.

{\small {\ }}
\begin{figure}
\begin{center}
\vspace{5mm} {\small
\begin{picture}(400,250)(0,-10)
\thicklines \Photon(40,250)(100,200)5 5 \Photon(40,50)(100,100)5 5
\ArrowLine(100,100)(100,200) \ArrowLine(250,100)(100,100)
\ArrowLine(250,200)(250,100) \ArrowLine(100,200)(250,200)
\Photon(250,200)(310,250)5 5 \Photon(250,100)(310,50)5 5
\Photon(170,200)(250,150)5 5
\end{picture}}
\end{center}
\caption{Pentabox diagram with six massive propagators. We
consider the case of equal masses and external particles on-shell.
At two-loop level it contributes to photon-photon scattering.}
\end{figure}
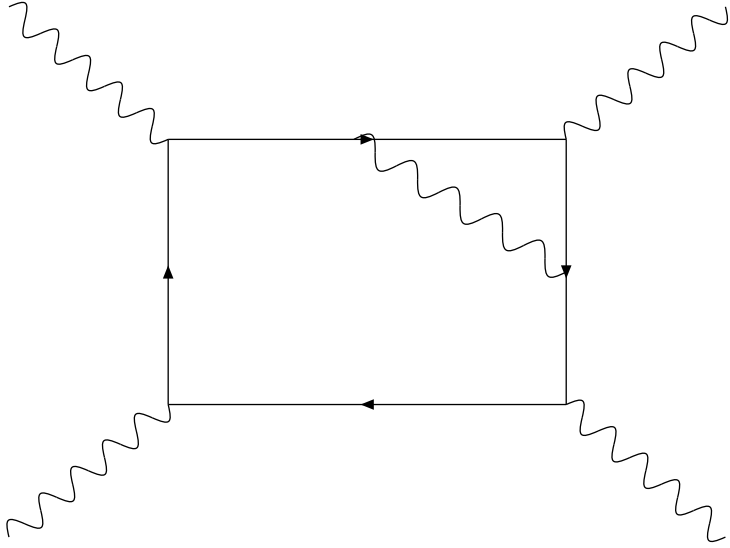

The simplest hypergeometric series representation is given by a
seven-fold summation,
\beq \label{7var}
({\cal MP})_7 &=& \pi^D\,(\mu^2)^{\s}\, P_7^{AC} \sum_{{\rm
all}=0}^\infty \frac{(-i|X_{123})(-j|X_{4567}) (-k|X_{14})
(-l|X_{257}) (-m|X_{367})(-n|X_{456})
(-r|X_{123})(-\s|X_{1234567}) }
{ (-m-\s|X_{1245}+2X_{367}) (-k-l|X_{12457})(D/2|X_{1234567})} \\
&\times&\frac{(-i-j-m-n-r-D/2|X_{12457}+2X_{36})(-k-l-m-D/2|X_7)(D/2+m|X_{1245})}{
(-i-j-n-r|2X_{123456}+X_7)X_1!...X_7!
}\left(\frac{s}{\mu^2}\right)^{X_{123}}
\left(\frac{t}{\mu^2}\right)^{X_{4567}},\nonumber
\eeq
where
\be P_7^{AC} = (D/2|m)(-\s|-m)(-i-j-n-r|-m-D/2)(-k-l|-m-D/2).
\ee

As a sample numerical calculation we give in the Table 1 below an expansion in the $\epsilon=2-D/2$ parameter.


\begin{table}
\begin{center} \caption{Column "terms" indicates where we
truncated all the series. Observe that even for a few terms in the
series that were summed we get a good precision.The result is of the
form $A+B\epsilon+C\epsilon^2$ where $A,B,C$ are given in the table.}
\begin{tabular}{lddd}
\hline Terms & $\epsilon^0$ & $\epsilon^1$& $\epsilon^2$ \\%
\hline 0 & 0.167027110 & -0.012614420 & 0.3914896\\%
\hline 1 & 0.167553265 & -0.012917188 & 0.3927286\\%
\hline 2 & 0.167558252 & -0.012920469 & 0.3927776\\%
\hline 3 & 0.167558333 & -0.012920522 & 0.3927779\\%
\hline 4 & 0.167558335 & -0.012920523 & 0.3927778\\%
\hline \end{tabular}
\end{center}
\end{table}

Hypergeometric series with ten summation indices, nine ``variables''
also occur,

\beq
({\cal MP})_9 &=& \pi^D\,t^\s\,P_9^{AC}\,\sum_{{\rm all}=0}^\infty
\frac{(-i|X_{123}+W_1)(-k|W_3+X_1+Z_1)(-n|W_5)(-r|W_6+X_{123})
(-\s|W_{123456}+X_{123})}{(1+j-\s|X_{123}+W_{13456})}\nonumber\\
&\times&\frac{(1-k-l-m-D|W_{34}-X_{123})(k+m+D/2|X_2-W_3-Z_1)
(-k-l-m-D/2|W_{34}+Z_1)}{(1-k-l-D/2|W_{34}-X_3)}\\
&\times &\frac{(i+j+r+D/2|-W_{126}-X_{123}+Z_1)
(1-\s-D/2|W_{123456})}{(-k-l-m-n-D/2|W_{345}+Z_1)}\frac{(s/t)^{X_{123}}\:(\mu^2/t)^{W_{123456}}}{X_1!...X_3!W_1!...W_6!Z_1!}\;,\nonumber
\eeq
where
\beq
P_9^{AC} &=& (-k-l-m-n-D/2|n)(\s+D/2|-2\s-D/2)
(k+l+m+D|-m-D/2)\nonumber\\
&\times &(-j|\s)(-l|k+l+m+D/2)(-m|i+j+m+r+D/2).
\eeq

Observe that there is a tenth series with unity argument, which is
not summable since it is a $_3F_2(a,b,c;e,f|1)$.

Finally the last result we present for the massive pentabox is a
11-fold series,

\beq
({\cal MP})_{11} &=& \pi^D\;(\mu^2)^{i+j+r+D/2}\:t^{k+l+m+n+D/2}\;P_{11}^{AC}\nonumber \\
&\times&\sum_{{\rm all}=0}^\infty
\frac{(-i|X_{123})(-k|X_1+W_3+Y_7+Z_1)(-n|W_5+Y_{789})(-r|X_{123})
(k+l+D/2|X_3+Y_9-W_{34})}{(1+j-k-l-m-n-D/2|W_{345}
+Y_{789}+Z_{12})}\nonumber\\ 
&\times &\frac{(1-k-l-m-n-D|-X_{123}+Y_{789}+Z_{12}+W_{345})
(1-k-l-m-D|W_{34}-X_{123}-Y_{789})}{(1-k-l-m-n-D|W_{345}-X_{123})
(1-k-m-D/2|W_3-X_2-Y_8+Z_1)}\nonumber \\
&\times &
\frac{(-i-j-r-D/2|X_{123}-Y_{789}-Z_{12})(-k-l-m-D/2|W_{34}+Z_{12})}
{(-i-j+k+l+m+n-r+D/2|-W_{345}+2X_{123}-Y_{789}-Z_{12})}
\nonumber\\
&\times& \frac{(s/\mu^2)^{X_{123}}\,(\mu^2/t)^{W_{345}+Y_{789}+Z_{12}}}
{X_1!X_2!X_3!Y_7!Y_8!Y_9!Z_1!Z_2!W_3!W_4!W_5!}\,,
\eeq
where
\beq
P_{11}^{AC} &=& (-i-j+k+l+m+n-r+D/2|i+r)(k+l+m+D|-m-D/2)\nonumber\\
&\times &(-j|-i-r-D/2)(-l|-k-m-D/2)(-m|k+2m+D/2)\,.
\eeq

It is important to note that all the series are valid within their
regions of convergence\cite{luke}, e.g., in the first one (\ref{7var})
one must have $|s/\mu^2|<1$ and $|t/\mu^2|<1$, and so on for others.

We mention that there are also other hypergeometric series, for instance,

\be
({\cal MP})_7 = (\mu^2)^{\s-n}t^n\sum_{{\rm all}=0}^\infty
\left(\frac{s}{\mu^2}\right)^{X_{abc}}
\left(\frac{\mu^2}{t}\right)^{X_{efgh}}\{(...|...)\}_1 +
(\mu^2)^{\s-j}t^j\sum_{{\rm all}=0}^\infty
\left(\frac{s}{\mu^2}\right)^{X_{abc}}\left(\frac{\mu^2}{t}\right)^{X_{efgh}}\{(...|...)\}_2
, \ee
and also 10-fold series, with 8 ``variables'', which has poles
in the exponents of propagators, like $\G(i-r)$,

\be
({\cal MP})_8 =\G(i-r)\G(...) \sum_{{\rm all}=0}^\infty
\left(\frac{t}{\mu^2}\right)^{X_{abcd}}\left(\frac{\mu^2}{s}\right)^{X_{efgh}}
+ \G(r-i)\G(...) \sum_{{\rm all}=0}^\infty
\left(\frac{t}{\mu^2}\right)^{X_{abcd}}\left(\frac{\mu^2}{s}
\right)^{X_{efgh}}, \ee
which is regularized introducing\cite{stand,box} for example,
$i=-1+\delta$ and then expanding around $\delta=0$.

Among all the possible (more than a million) hypergeometric series
representations for the integral in question there must be many
relations among them by \ac{} (directly or indirectly). Yet, the study
of them all is such a formidable task that for practical purposes it
is virtually impossible to even think of doing it thoroughly, so we
can only conjecture that the entire set of hypergeometric series
representations cover all the kinematical manifold. Also, these \ac{}
formulas can in principle be obtained from them.

It is worth remembering some important points in the complex analysis
theory\cite{feshbach}. The reader can observe that some of the results
have poles of different order than others, i.e., some of them are like
$1/\epsilon^a$ and other(s) like $1/\epsilon^b$, with $a\neq b$.  In
our previous work on box integrals for photon-photon scattering the
same occured\cite{box}. In that case we had a branch cut at $s=4m^2$,
and some hypergeometric functions had a region of convergence which
allowed us to study such integrals in that point. On the other hand,
such hypergeometric series had a direct \ac{} into another one which
did not allow us to consider $s=4m^2$. The answer is in
\cite{feshbach}: when we carry out an \ac{} and in the process cross a
branch cut, this \ac{} is not unique and poles do appear. In that
case, there were simple and double poles, and we had two possible
cases: $F_3\rightarrow \Sigma F_2$ and $F_3\rightarrow H_2$. In our
present case, $({\cal MP})_{11}$ and $({\cal MP})_9$ have third order
poles, and $({\cal MP})_7$ is finite. It depends on the kinematical
region we wish to study the pertinent integral.

Back in the 50's and 60's \cite{landshoff} a lot of research was
done in order to study singularities of Feynman integrals. One of
the results we borrow from them is that a graph like the one
considered in the present section can not have singularities in
the physical sheet. This is a well-known theorem due to Eden
(1952). So, in a full calculation of photon-photon scattering the
poles must cancel out.

In order to extract specific pole structures of these integrals, we
can proceed just like in our previous works on the
subject\cite{stand}. Expand gamma functions around $\epsilon=0$ and
use Taylor expansion in the hypergeometric series, so that we are left
with parametric derivatives of \hf{}s.  The reader can see detailed
calculations in the above referred paper; also in the appendix of
\cite{coulomb} and in the next section.

\subsection{Non-Planar Double-Box integral}

Now we turn to massless non-planar double-box with six
propagators, namely,
\be
{\cal NP}_6 = \int d^D\! q\; d^D\! r\;
(q^2)^i (q-p_3)^{2j} (q+r)^{2k} (q+r+p_2)^{2l} (r^2)^m
(r-p_4)^{2n},
\ee
which represents the diagram of Figure 4.  This diagram was also
studied by Smirnov and Veretin\cite{veretin} that presented an
explicit result in the case where all exponents were equal to minus
one. On the other hand, we will write down results for arbitrary
values of them.

The generating functional

\beq
G_{{\cal NP}_6} &=& \int d^D\! q\; d^D\! r\;
\exp{\left\{-\alpha q^2 -\beta (q-p_3)^2 -\gamma (q+r)^2 -\theta
(q+r+p_2)^2 -\phi r^2 -\omega (r-p_4)^2\right\}} \nonumber\\ &=&
\left(\frac{\pi^2}{\zeta}\right)^{D/2}
\exp{\left[-\frac{1}{\zeta}\left( \beta \gamma\omega s +
\alpha\theta\omega t + \beta\theta\phi u \right)\right]},
\eeq
can be integrated out without difficulty. We define
$\lambda_3=\alpha+\beta+\gamma+\theta$ and $\zeta=\alpha\gamma+
\alpha\theta +\beta\gamma+\beta\theta+\lambda_3(\phi+\omega)$. It
is easy to see that Taylor expanding the above exponential will give
us three series, while multinomial expansion for $\zeta$ other twelve
series. The equations that form the system to be solved come from the
propagators, six altogether, and an additional constraint originates
from the multinomial expansion.

So, at the end of the day we are left with $(15-7)=8$-fold series.
Their variables are either,
\be
\frac{t}{s},\quad \frac{s}{t},\quad \frac{t}{u},
\quad \frac{u}{t},\quad \frac{u}{s},
\quad \frac{s}{u}, \ee
and/or unity. The simplest \hsr{} for ${\cal NP}_6$ are double series,

\be {\cal NP}_6 = \pi^D\,s^{\s'}\,\G_{{\sc
NP}}^{AC}\sum_{X_2,X_3=0}^\infty \frac{(-i|X_2)(-m|X_3)(-\s'|X_{23})
(-l|X_{23})}{(1+k-\s'|X_{23})(1+n-\s'|X_3) (1+j-\s'|X_2)}\,
\frac {(-t/s)^{X_2}\,(u/s)^{X_3}}{X_2!\,X_3!}\,,
\label{np6-dupla} \ee
where $\s'=i+j+k+l+m+n+D$, is the sum of exponents and dimension and

\beq
\G_{{\sc NP}}^{AC} &=& (-j|\s')(-k|\s')(-n|\s')(\s'+D/2|-2\s'-D/2)
(i+j+m+n+D|-i-j-D/2) \nonumber\\
&\times & (k+l+m+n+D|-m-n-D/2)(i+j+k+l+D|-k-l-D/2). \eeq

The above hypergeometric series reduces to an Appel $F_2$ function
in the special case where $k=-1$,

\be
{\cal NP}_6 = \pi^D\, s^{\s'} \G_{{\sc NP}}^{AC}(k=-1)\; F_2
 \left.\left( \matrix{ -l, -i, -m \cr 1+j-\s', 1+n-\s'
 }\right| \frac{-t}{s},\frac{u}{s}\right).\ee

If we take all the exponents to be equal to minus one, we have

\beq {\cal NP}_6 &=& \pi^D\,
s^{D-6}\frac{\G(6-D)\G^3(D/2-2)\G^3(D-5)}{\G(3D/2-6)\G^3(D-4)}
\sum_{X_2,X_3=0}^\infty
\frac{(1|X_2)(1|X_3)(1|X_{23})}{(6-D|X_3)(6-D|X_2)}\,
\frac{(-t/s)^{X_2}\,(u/s)^{X_3}}{X_2!\,X_3!} \label{np6-1} \\
&=&\label{resultado} \pi^D\, s^{D-6}\, \frac{\G(6-D)
\G^3(D/2-2)\G^3(D-5)}{\G(3D/2-6)\G^3(D-4)}\, F_2
 \left.\left( \matrix{ 1, 1, 1 \cr 6-D, 6-D }\right|
 \frac{-t}{s},\frac{u}{s}\right).
\eeq

This series is the Appel\cite{luke} $F_2$ \hf{} which converges
when $|t/s|<1$, $|u/s|<1$ and $|t/s|+|u/s|<1$. Note that the above
result has a double pole, $1/\epsilon^2$, just like in
Tausk's\cite{tausk} {\bf and in Smirnov and Veretin's \cite{veretin}.}
works.

One could now ask: Would the divergent pieces come from a few terms in
the series or from all of them? Clearly, in the present case,
divergent factors which generate double and simple poles come from the
gamma functions. To write down these poles explictly we have to Taylor
expand also the \hf{} $F_2$, then we have around $\epsilon=0$
($D=4-2\epsilon$), \beq {\cal NP}_6
\label{resultado-f2} &=& \pi^D\,
s^{D-6}\,\frac{\G(6-D)\G^3(D/2-2)\G^3(D-5)}{\G(3D/2-6)\G^3(D-4)}\; F_2
\left.\left( \matrix{ 1, 1, 1 \cr 6-D, 6-D }\right|
-\frac{t}{s},\frac{u}{s}\right) \\ &=& \pi^D\, T \,s^{D-6}\,
\left[-\frac{3}{\epsilon^2} +12 +{\cal O}(\epsilon) \right]\nonumber\\
&\times& \left\{ F_2 \left.\left( \matrix{ 1, 1, 1 \cr 2, 2 }\right|
-\frac{t}{s},\frac{u}{s}\right)+ 2\epsilon (\partial_\gamma +
\partial_{\gamma'}) F_2\right. +\left. 2\epsilon^2 \left[
\partial_\gamma^2 + \partial_{\gamma'}^2
+4\partial_\gamma\partial_{\gamma'}\right]F_2+{\cal
O}(\epsilon^3)\frac{}{} \right\},\nonumber \eeq where $\gamma_E$ is
the Euler's constant and
$$ \partial_\gamma F_2 = \frac{\partial}{\partial\gamma}F_2
\left.\left.\left( \matrix{ \alpha, \beta, \beta' \cr \gamma,
\gamma'}\right| -\frac{t}{s},\frac{u}{s}\right)
\right|_{\alpha=\beta=\beta'=1,\; \gamma=\gamma'=2}, $$ are called
parametric derivatives of \hf{}s and can be calculated using
Davydychev's \cite{stand} approach.($T$ is given below.)

In order to rewrite the above result as polilogarithm functions, such
as ${\rm Li}_2$, ${\rm Li}_3$ and usual logarithms we have to use an
integral representation for $F_2$, \be
\frac{\G(\beta)\G(\beta')\G(\gamma-\beta)\G(\gamma'-\beta')}{\G(\gamma)\G(\gamma')}
F_2 \left.\left[ \matrix{ \alpha; \beta, \beta' \cr \gamma ,
\gamma'}\right|P_1, P_2\right] = \int_0^1 dx_1 dx_2
\frac{x_1^{\beta-1} x_2^{\beta'-1} (1-x_1)^{ \gamma-\beta-1} (1-x_2)^{
\gamma'-\beta'-1} }{(1-x_1P_1 -x_2P_2)^\alpha}, \ee where
$x_1+x_2=1$. This task in not an easy one at all, since the second
derivatives give a very cumbersome result and direct comparison
between our result and Tausk's \cite{tausk} is not possible (we were
not able to show that both results are equivalent analytically).

However, the important special case of forward scattering can not be
read off directly from Tausk's result. On the contrary, making $t=0$
or $u=0$ is extremely simple in our eq.(\ref{resultado}).  Let $u=0$,
then one of the series (in $X_3$ index) does not contribute and \be
F_2 \left.\left( \matrix{ 1, 1, 1 \cr 6-D, 6-D }\right|
-\frac{t}{s},0\right) =\; _2F_1 \left.\left( \matrix{ 1, 1 \cr
6-D}\right|-\frac{t}{s},0\right) = \frac{\G(6-D) \G(4-D)}{\G^2(5-D)},
\label{f21} \ee since $s+t+u=0$ we were able to sum up the series
$_2F_1$, so $\partial/\partial\gamma$ can be rewritten in terms of
$\partial/\partial\epsilon$ and parametric derivatives are easily
calculated from eq.(\ref{f21}). We quote the result up to order
$\epsilon^0$,

\beq \left.{\cal NP}_6\right|_{u=0} &=& \pi^D s^{-2-2\epsilon}T
\left\{ -\frac{48}{\epsilon^4} + \frac{201}{2\epsilon^3} -
\frac{204}{\epsilon^2} + \frac{8}{\epsilon} \left[83+ 16\psi^{''}(1) -
16\psi^{''}(2)\right]\right.\nonumber\\ && \left. +4\left[
-134+35\psi^{''}(1) -35\psi^{''}(2)\right] + {\cal
O}(\epsilon)\frac{}{}\right\}, \eeq
where $\psi(z)\G(z)=\G'(z)$, $\psi'(z)$, $\psi''(z)$ their derivatives
and
$$ T = \frac{\G(5-D)\G^3(D/2-1)}{\G(3D/2-5)(5-D)}, $$ which in fact
can be calculated to arbitrary order, since we start from an exact
result.

Observe that the result has poles of higher orders, namely
$\epsilon^{-4}$ and $\epsilon^{-3}$, which come from the fact we
have taken $u=0$, which also mean that the result has a branch
point in $u=0$, a well-known fact from the theory of
hypergeometric functions.

Of course, the $7\times 15$ system of linear algebraic equations
defined by this integral, the $7\times 7$ solvable subsystems are
divided into the following categories: among the grand total of
$6,435$ possible solutions, $3,519$ have no solution at all. Among
the remaining $2,916$ relevant solutions, \ndim{} provides other
kind of series, such as 7-fold series and 5-fold series. And all
of them have symmetries among $s,t$ and $u$, namely,

\be (p_3\leftrightarrow p_4,\; j\lra n, \; i\lra m, \; t\lra u)\,,
\qquad (p_2\leftrightarrow p_3,\; l\lra n, \; k\lra m, \; t\lra
s)\,,\qquad (p_2\leftrightarrow p_4,\; j\lra l, \; i\lra k, \; s\lra
u)\,, \label{simetrias} \ee

So, for each \hsr{} provided by \ndim{} there are other two, also
originated from the system of algebraic equations, which represent
the same integral and can be transformed into the first using
(\ref{simetrias}). This is the case of (\ref{np6-dupla}), and it
is stated also by Tausk\cite{tausk}, i.e., the diagram is
completely symmetric (through \ac{}) under external legs
permutations.

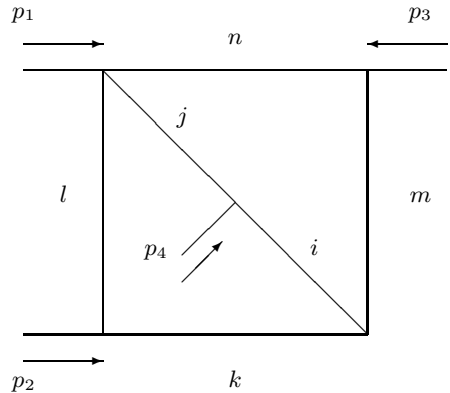
\begin{figure}
\begin{center}
\begin{picture}(600,200)(0,150)
\vspace{40mm} \thinlines \put(150,200){\line(1,0){130}}
\put(150,190){\vector(1,0){30}} \put(150,300){\line(1,0){160}}
\put(150,310){\vector(1,0){30}} \put(180,200){\line(0,1){100}}
\put(280,300){\line(0,-1){100}} \put(280,200){\line(-1,1){100}}
\put(230,250){\line(-1,-1){20}} \put(310,310){\vector(-1,0){30}}
\put(210,220){\vector(1,1){15}}

{\small \put(230,180){\makebox(0,0)[b]{$k$}}
\put(300,320){\makebox(0,0)[b]{$p_3$}}
\put(150,320){\makebox(0,0)[b]{$p_1$}}
\put(150,180){\makebox(0,0)[b]{$p_2$}}
\put(200,230){\makebox(0,0)[b]{$p_4$}}
\put(230,310){\makebox(0,0)[b]{$n$}}
\put(260,230){\makebox(0,0)[b]{$i$}}
\put(165,250){\makebox(0,0)[b]{$l$}}
\put(210,280){\makebox(0,0)[b]{$j$}}
\put(300,250){\makebox(0,0)[b]{$m$}}}
\end{picture}\caption{Scalar massless non-planar double-box with six propagators. The labels in
the internal lines represent the exponents of propagators, see
(\ref{np6}). }
\end{center}\end{figure}

Despite the complicated form of non-planar double-box with six
propagators, the result we obtained is very simple, a double
hypergeometric series, which in the special case of $k=-1$ reduces
to $F_2$ Appel function.  Tausk presented a result for the same
graph in terms of di- and trilogarithms, in the special case where
all exponents are equal to minus one.  If we recall \cite{stand},
Davydychev calculated a scalar integral for photon-photon
scattering and transformed the four $F_2$ Appel \hf s into
dilogarithms. The same occurs here, in one hand di and
trilogarithms and in the other double hypergeometric series,
namely $F_2$.

\section{Conclusion}

In this work we considered covariant gauge scalar pentabox and
non-planar double-box integrals. In the former we considered two
cases: all virtual particles being massless, and in the other, six
of them having the same mass $\mu$ while the seventh is massless.
The latter was calculated in the massless case and arbitrary
exponents of propagators, a result which was missing in the
literature. No reduction formulas, i.e., rules to connect a given
diagram with simpler ones, were used. The results are given in
terms of hypergeometric series and in different kinematical
regions.

\acknowledgments{ AGMS gratefully acknowledges FAPESP (Funda\c
c\~ao de Amparo \`a Pesquisa de S\~ao Paulo) for financial
support.}

\end{document}